\documentclass[doublecol]{epl2} 
\usepackage{amsmath,amssymb,mathrsfs}

\title{Experimental study of work exchange with a granular gas: \\the viewpoint of the Fluctuation Theorem.}
\shorttitle{Statistics of work exchange with a granular gas.} 

\author{Antoine Naert.}
\shortauthor{Antoine Naert.}

\institute{                    
Laboratoire de Physique de l'\'Ecole Normale Sup\'erieure de Lyon, Universit\'e de Lyon, CNRS UMR 5672, \\46 All\'ee d'Italie, 69364 Lyon cedex 7, France.\\
}

\abstract{
This article reports on an experimental study of the fluctuations of energy flux between a granular gas and a small driven harmonic oscillator. The DC-motor driving this system is used simultaneously as actuator and probe. The statistics of work fluctuations at controlled forcing, between the motor and the gas are examined from the viewpoint of the Fluctuation Theorem. A characteristic energy $E_c$ of the granular gas, is obtained from this relation between the probabilities of an event and its reversal.}

\begin{document}
\maketitle

\section{Introduction}
\label{intro}
Often, the principle of probing is to measure the response of a system to an excitation imposed from outside. This excitation must be small enough not to perturb the state of the system under scrutiny. Together with the Fluctuation Theorem (FT), this principle is used in the present study to probe the disordered motion of a granular gas.\\ 
In addition to the power needed to keep the granular gas in a Non Equilibrium Steady State (NESS), ({\it i.e.} off-setting the mean dissipation), a small probing power is imposed. The response to this excitation is characteristic of the granular gas disordered {\it state}. More precisely here, it is the product of the forcing perturbation and the velocity response, {\it i.e.} the power perturbation, that carries information on the system: its statistical properties will be considered in order to characterise the disordered granular gas NESS. \\
The measurement (excitation-response), is performed by means of a driven harmonic oscillator {\it coupled} to the granular gas. The Fluctuation Theorem (FT) is used to compare the work given and received from the granular gas by the harmonic oscillator. It is the Steady State Fluctuation Theorem (SSFT) that is considered all along this article, a relation holding in the limit of asymptotically large times. No further mention of this distinction will be given later.  \\
The power provided (work flux) can be related to the rate of entropy production $\sigma$. The FT states a relation between the probability of events of positive $\sigma$, and that of equal but negative entropy production rate. It states that the ratio of these probabilities simply  increases as the exponential of $\sigma$ \cite{ft}. It therefore quantifies the failure of detailed balance, for a large number of degrees of freedom chaotic dynamical system. \\
This experimental study applies an original and very simple probing principle, to show that FT seems to holds in a stationary granular gas, and measure a {\it characteristic energy} $E_c$. It is a contribution to the very few experimental studies of the different forms of the FT. (For a nice review, see \cite{ciliberto2010}.)\\
Rather few experiments invoking the FT have been actually done in granular gases particularly. Most studies  are numerical \cite{aumaitre2004}, or theoretical \cite{puglisi2005}. Strictly speaking, the only experimental study of a granular gas with the FT is that of Feitosa and Menon \cite{feitosa2004}. In a 2D vibrated granular material, they measured by video tracking the momentum flux of particles in and out of a sub-volume, and used the FT to define an effective temperature of the medium. The interpretation of these results is difficult, and still questioned \cite{puglisi2005}. \\
In a distinct context, D'Anna {\it et. al.} performed rheology in dense granular fluids using a driven torsion pendulum. Thanks to an out-of-equilibrium extension of the Fluctuation-Dissipation Theorem, they defined an effective temperature \cite{danna2003}.\\
The present article describes an original study of a dilute 3D granular gas, also using a driven torsion pendulum. In a simpler but reliable manner, it show how to measure an energy $E_c$, characteristic of the particles disordered motion, thanks to the FT. \\
The equation of motion of the harmonic oscillator writes as the following:  
\begin{equation}
\label{ho}
M\ddot{\theta}+\gamma \dot{\theta}+k\theta=\Gamma(t)+\eta(t), 
\end{equation}
where $\theta$ is the angle of torsion of the pendulum, and dots stand for time derivatives. $M$, $\gamma$, $k$ are respectively moment of inertia, viscous friction coefficient, and spring constant. A sine torque $\Gamma(t)$ is imposed from outside. The last term $\eta(t)$ represents the coupling with the NESS granular gas {\it heat bath}. In the framework of Langevin equation description, it represents the momentum transfer rate at each shock from the beads. \\
When the FT is expressed in terms of the work variation rate $\dot w(t)=\Gamma\,\dot{\theta}$, the mechanical power transmitted during a time-lag $\tau$ to the gas is expressed as: $\dot w_\tau(t)=\frac{1}{\tau}\int_{\tau}\dot w(t') dt'$. It relates the probabilities of giving a power $\dot w_\tau$ to the gas and the probability of receiving the same amount from it. It states that this ratio increases exponentially with the coarse-grained work $\tau \dot w_\tau$:  
\begin{equation}
\label{ft}
\frac{\Pi(\dot w_\tau)}{\Pi(-\dot w_\tau)} = e^{\tau \dot w_{\tau}/E_c},
\end{equation}
for asymptotically large times $\tau$. $\Pi$ is the probability, and the coefficient $E_c$ is a {\it characteristic energy} (possibly linked to the mean kinetic energy of the gas). \\
The principle of the measurement and the experimental set-up are described in the next section. The results obtained are detailed in the following section. The last section is devoted to a brief discussion of these results.

\section{Measurement's principle}
\label{experiment}
\begin{figure}[!h]
\begin{center}
\onefigure[width=8.7cm]{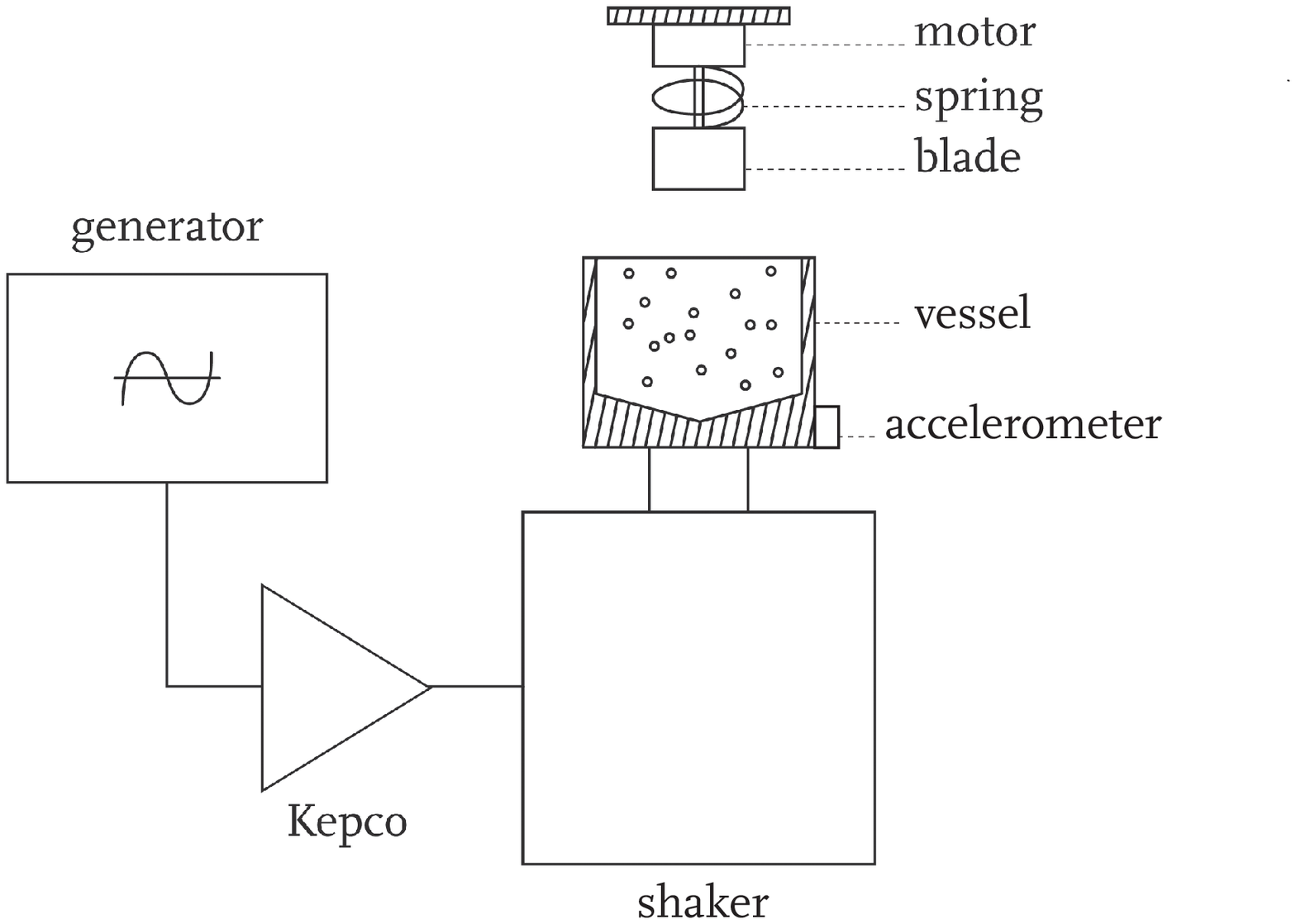}
\end{center}
\caption{The mechanical system is composed of a vibrating vessel with its driving, and the probing harmonic oscillator, pulled out for clarity.}
\label{fig.1}
\end{figure}

Fig.~\ref{fig.1} sketches the experimental set-up. The granular gas is composed of a few hundreds of $3\,$mm diameter stainless steel beads, contained in a vibrating vessel. 

The vessel is aluminum made, $5\,$cm diameter, $6\,$cm deep, its inside bottom is slightly cone-shaped to favour horizontal momentum transfer (angle=$10^\circ$). (See fig.~\ref{fig.1}.) The electromechanical shaker is driven by a sine generator, via a Kepco current amplifier. An accelerometer fixed on the vibrating vessel measures the vertical acceleration: between $41\,$ms$^{-2}$ and $60\,$ms$^{-2}$, at a frequency $f_{\rm{exc.}}=40\,$Hz. A small DC-motor is simultaneously used as actuator and sensor. It is a regular permanent magnet, brushed DC-motor, of relatively small size ($25\,$mm diameter). A plastic blade of approximately $20$ x $20\,$mm is fixed on the axis of the motor. A torsion spring is used to produce an elastic force on the motor axis. This system motor + blade + spring forms a harmonic oscillator. Its resonance frequency is a few hundreds of Hz, in any case higher than any frequency of the signal of interest. The motor is fixed on a cover closing the vessel, which prevents the beads from hopping off the vessel.\\
The principle of the measurement is the following. A DC-motor can be used reversibly as a generator. As a generator, the induced voltage is proportional to the angular velocity: $e=\alpha \dot \theta$. As a motor, the torque is proportional to the current supplied $I$: $\Gamma = \alpha\,I$. Notice that both relations involve the same proportionality coefficient, which depends on the physical parameters of the motor itself. The mechanical work produced by the motor against the granular gas per unit time is simply: $\dot w = \Gamma \, \dot \theta = e\,I$.

\begin{figure}[h!]
\begin{center}
\onefigure[width=8.5cm]{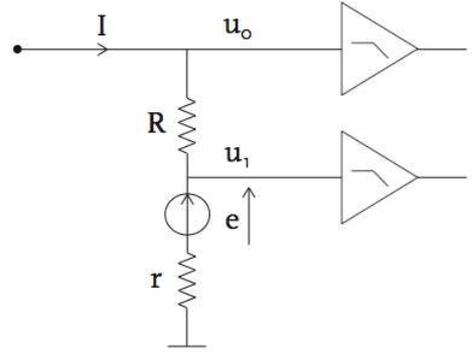}
\end{center}
\caption{The electrical sketch of the motor's command, and measurement set-up, that is to say the amplifiers-filters stage. The motor is represented by the induced voltage 'source' $e$ and its internal resistance $r$.}
\label{fig.2}
\end{figure}

\begin{figure}[!h]
\onefigure[width=8.8cm, height=6.9cm]{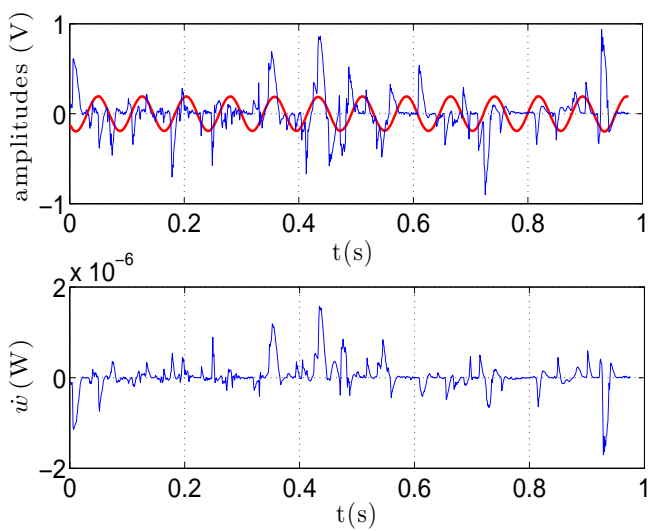}
\caption{Top: time-series of the imposed sine current ($I \simeq 200 \mu A$), and the induced voltage. For homogeneity and comparison, both have been plotted in Volts: $R\;I(t)$ (bold), and for readability, $e$ has been enlarged by a factor $100$ (thin). Bottom: in the same time-scale, the corresponding energy flux $\dot w(t)$ is plotted. The mean power given is $< \dot w >\simeq 7.3\;nW$. The acceleration is $a \simeq 56\; $ms$^{-2}$. } 
\label{fig.3}
\end{figure}

The motors command and measurement set-up are sketched in fig.~\ref{fig.2}. The harmonic oscillator is driven by an AC current, supplied through a $R=1\,k\Omega$ resistor by a sine voltage generator at $f_e=13\,$Hz frequency. 
As $R$ is large enough, the motor is driven by a periodic current (see time-series in fig.~\ref{fig.3}). At such a low frequency, iron losses in the motor are negligible. The voltages $u_0$ and $u_1$ are recorded by a $16$ bits simultaneous acquisition board at frequency $f_s=1024\,$Hz. The amplificators adjusts the level of the signals to that of the A/N converter, and anti-alias filters at $512\,$Hz. The results reported here are obtained from one hour-long recordings. The current $I(t)=(u_0-u_1)/R$ and the induction voltage $e(t)=u_1-r\,I(t)$ are measured, $r$ being the internal resistance of the motor ($r\simeq5.8\,\Omega$). \\
The power $\dot w(t)=\Gamma \, \dot \theta = e\,I$ produced by the motor at $13\,$Hz frequency is completely transferred to the granulafr gas. 
Although $I(t)$ is a sine, $e(t)$ fluctuates strongly because of the beads collisions on the blade. The power $\dot w = e\,I$ shows large fluctuations, negative whenever $I$ and $e$ have opposite sign (see fig.~\ref{fig.3}). In such case, the reservoir is giving work to the blade-motor system. The fluctuations of power are widely distributed positively and negatively. 
\begin{figure}[!h]
\onefigure[width=8.8cm, height=5.7cm]{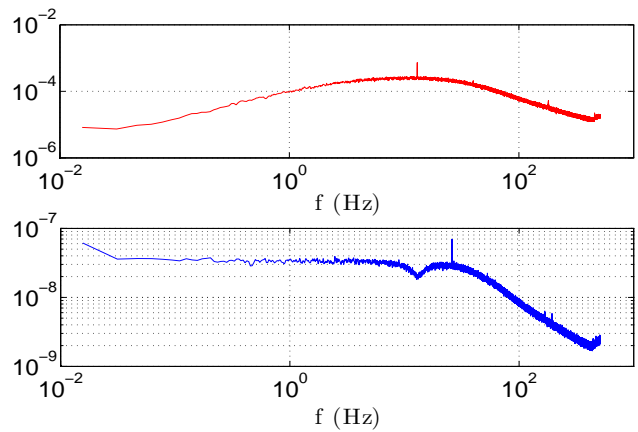}
\caption{For the same as the previous measurement, the power spectral densities of $e$ is plotted on the top, showing that the induced voltage (velocity) is broad-band noise. The excitation is visible at $f_e=13\,$Hz. The bottom plot shows the power spectral density of the probing power $\dot w$.}
\label{fig.4}
\end{figure}
\noindent
Looking carefully at the time series, one cannot see in the fluctuations of the voltage any trace of the sine excitation. One can see in fig.~\ref{fig.4} the power spectral densities of the induced voltage $e$. It is broad band noise, reflecting that the momentum transfer from the granular gas is disordered: no periodic contribution is visible except a small contribution at $13\;Hz$ coming from the current. 
\section{Results}
\label{results}
\begin{figure}[!h]
\begin{center}
\onefigure[width=8.8cm, height=5.8cm]{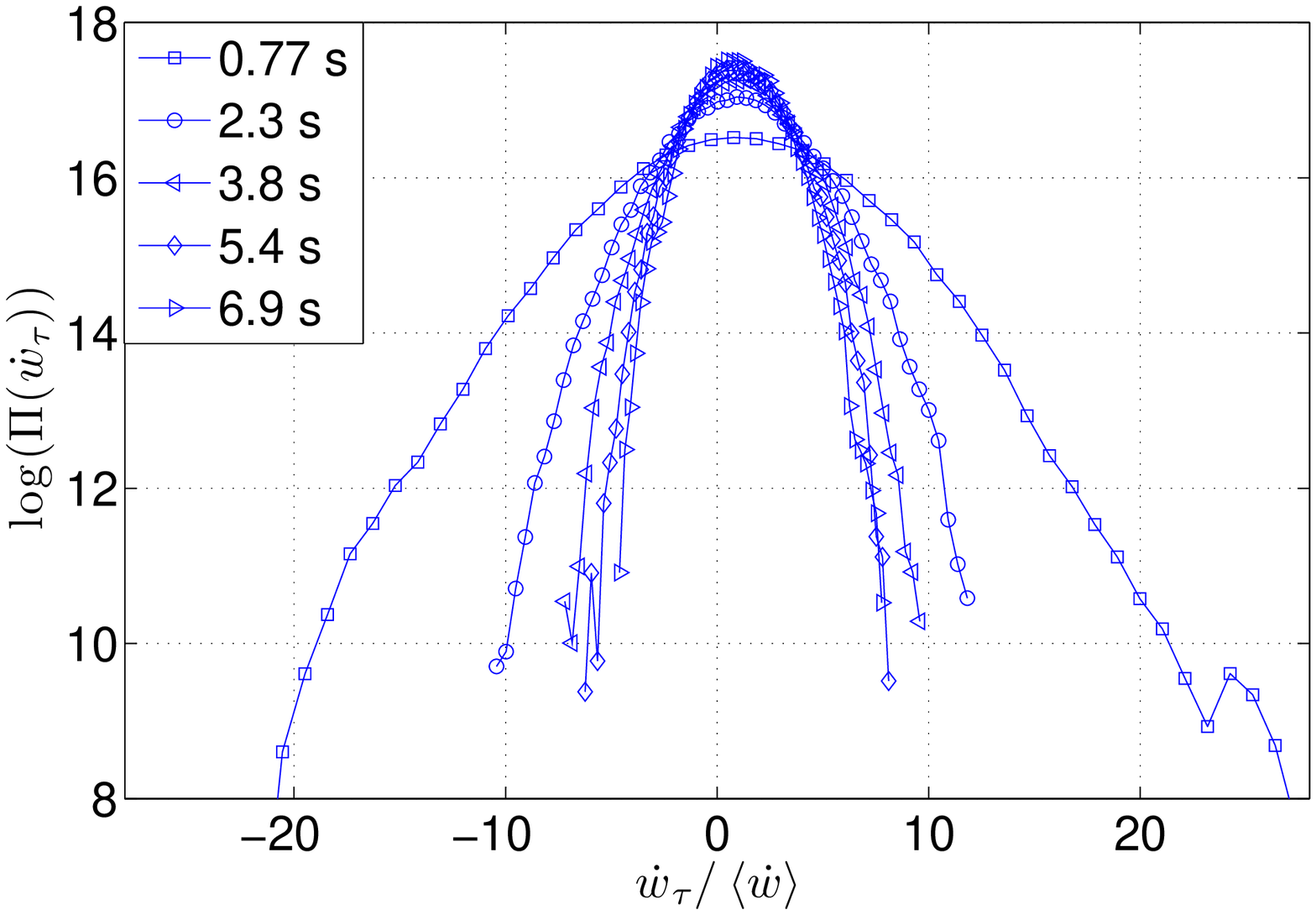}
\end{center}
\caption{Histograms of the power given to the gas for $I \simeq 200 \mu A$, time-averaged over several values of $\tau$ ($\Box$: 0.77\,s,  $\bigcirc$: 2.3\,s, $\lhd$: 3.8\,s, $\diamond$: 5.4\,s, $\rhd$: 6.9\,s.), {\it i.e.} $980$ to $8.9\;10^3\,\tau_c$, where $\tau_c \simeq 7.8\;10^{-4}\,s$. $\dot w_\tau $ is normalised by the mean value $<\dot w> \simeq 5.3\;10^{-9}\, W$.}
\label{fig.5}
\end{figure}

\begin{figure}[h!]
\begin{center}
\onefigure[width=8.8cm, height=5.8cm]{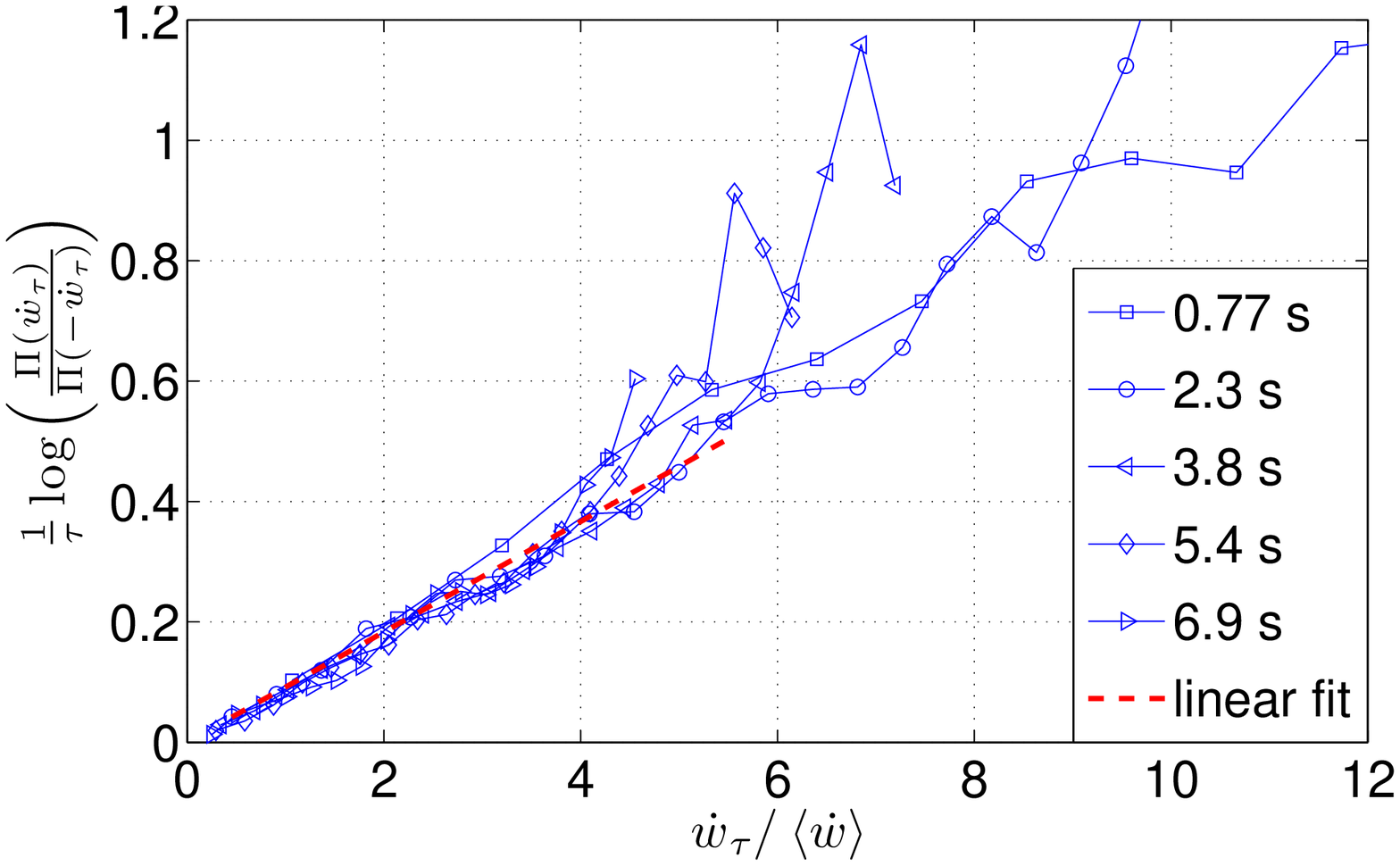}
\end{center}
\caption{The asymmetry function $\frac{1}{\tau}log\left(\frac{\Pi(\dot w_\tau)}{\Pi(-\dot w_\tau)}\right)$ is plotted against $\dot w_\tau/<\dot w_\tau>$, for the same 5 time-lags as on previous figure ($\Box$: 0.77\,s,  $\bigcirc$: 2.3\,s, $\lhd$: 3.8\,s, $\diamond$: 5.4\,s, $\rhd$: 6.9\,s, {\it i.e.} $980$ to $8.9\;10^3\,\tau_c$, where $\tau_c \simeq 7.8\;10^{-4}\,$s). ($I \simeq 200 \mu A$.) A linear fitting is performed, shown only after transient, for $\tau\sim2.3\,s$ (dashed line).}
\label{fig.6}
\end{figure}

Histograms of the power $\dot w_{\tau}$ transmitted to the gas over several time-lags $\tau$ are calculated: $\dot w_{\tau}(t) =\frac{1}{\tau}\int_{\tau} e(t') I(t') dt'$. Time-lags are chosen as integer multiples of the excitation period: $\tau=n/f_e$. Convenient values are chosen as $n=10, 30, 50, 70, 90$. Corresponding histograms are shown in fig.~\ref{fig.5}. They are calculated over 50 bins. The wider histograms correspond to the smallest $\tau$. 
Starting from those histograms, eq.~\ref{ft} is examined, plotting against $\dot w_\tau$ the so-called {\it asymmetry function}: $\frac{1}{\tau}log\left(\frac{\Pi(\dot w_\tau)}{\Pi(-\dot w_\tau)}\right)$. Here and in the following, the natural logarithm only is considered. One can see in fig.~\ref{fig.6} that these curves are approximately linear, at least for moderate $\dot w_\tau$ ($\tau$ not too small), validating experimentally the prediction of  FT. In some cases however, curves are observed to be slightly bent downward for large values of $\dot w_{\tau}$. This is due to $\tau$ being too small. \\
In order to test eq.~\ref{ft}, the slope of the asymmetry function is evaluated. To avoid the fitting being dominated by extreme but less relevant values, it is performed for each $\tau$ only over amplitudes  $\dot w_{\tau}$ less than half of the maxima. A ratio slightly different, like $1/3$ or $2/3$, gives the same results. With $1/2$, this procedure is stable for all experimental configurations considered here (change of $I$, and acceleration of the vessel).  

\begin{figure}[h!]
  \begin{center}
\onefigure[width=8.8cm, height=5.5cm]{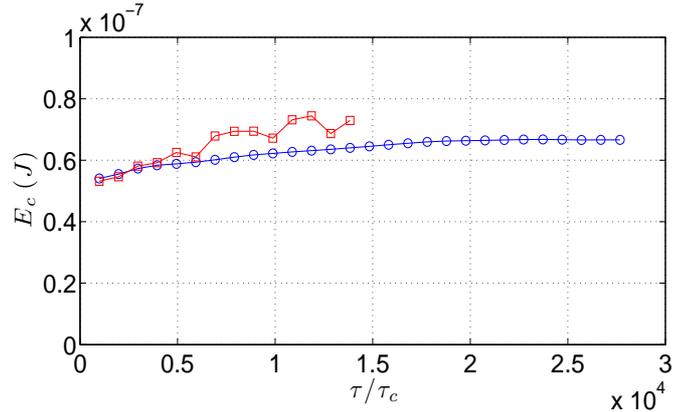}
  \end{center}
\caption{$E_c$ measured at $I \simeq 200 \mu A$, against $\tau/\tau_c$, $\tau_c$ being the correlation time of the process $\dot w$. The squares and circles refer to $E_c$ measured according respectively to the first and second method. $a=56\,$ms$^{-2}$.}
\label{fig.7}
\end{figure}

\noindent 
The fitting parameter is the slope $1/E_c$, where $E_c$ is called {\it characteristic energy}. It is evaluated for different values of the time-lag $\tau$. It tends to converge toward a limit value at large $\tau$ (see fig.~\ref{fig.6} and fig.~\ref{fig.7}). \\
For long times though, negative events become rare, causing statistical uncertainty on $E_c$. For this reason, the parameters must be chosen such that $< \dot w>$ is small, and fluctuation large. However, for short time-lags, as fluctuations are dominant, this asymmetry-based method becomes unreliable. The whole procedure relies on a delicate compromise. The fluctuation rate ${<\left(\Delta\dot w_\tau\right)^2>}^{1/2}/<\dot w_\tau>$ is a decisive parameter for the reliability of this measurement. In this study, it is between $20$ and $60$. It is a very useful feature of this experiment to adjust $<\dot w_\tau>$ at will. This is labelled as the first method to measure $E_c$.\\ 
It can be noticed on fig.~\ref{fig.3} that the histograms are close to Gaussian, when $\tau$ increases. This is expected in the limit of large $\tau$ from central limit theorem for variables such as $\dot w_\tau$, that result from a summation. It does not mean that $\dot w$ is Gaussian. Actually it is more like exponential, but the behavior at small $\tau$ are not of interest here. Additionally, it has been checked that the variance goes as $1/\tau$ as the mean is constant. \\
\indent
For a Gaussian variable, it can easily be proved that the FT takes the very simple form:  
\begin{equation}
\label{gaussian_approximation}
E_c=\frac{\tau}{2} \frac{< \dot w_\tau^2>}{< \dot w_\tau>}.
\end{equation}
Note that such a relation between the two sole moments of a Gaussian process is remarkable: the FT relates it by introducing the characteristic energy. \\
In addition to the first method (slope of the asymmetry function), the calculation of $E_c$ has been performed thanks to eq.~\ref{gaussian_approximation}. This second method to measure $E_c$, valid under Gaussian hypothesis, does not require negative events. \\
Those two methods are expected to give a close result, as $\dot w_\tau$ is close to Gaussian. 
Despite a large scattering of the slope measurements, fig.~\ref{fig.7} shows that they are consistent within a range of about $20 \%$, at least as far as negative events occur in large enough number to measure $E_c$. A discrepancy of this order of magnitude has been observed in all measurements. The process $\dot w_\tau$ is near Gaussian, but seems distinct at the present stage. \\
Although the first method gives results more scattered than the second one, it will be preferred in the following as it is more general: no Gaussian assumption is needed. 

\begin{figure}[h!]
\begin{center}
\onefigure[width=8.8cm, height=5.4cm]{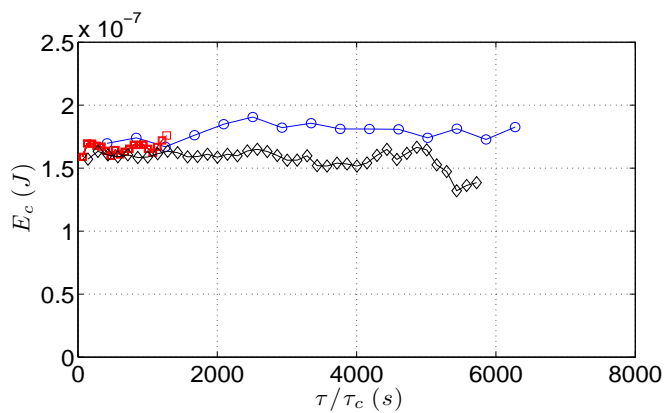}
\end{center}
\caption{$E_c$ against $\tau/\tau_c$, for several values of the excitation current I ($\Box:1\,$mA$,\;\Diamond$: $700\,\mu$A, and $\bigcirc$: $400\,\mu$A), but the same acceleration: $a=41\,$ms$^{-2}$. }
\label{fig.8}
\end{figure}

\noindent
It is interesting to compute the correlation time $\tau_c$ of $\dot w$. The loss of correlation is rather fast, as the level of the periodic excitation is not much larger than the noise of the granular gas (see fig.~\ref{fig.4}, top). In all the measurements shown, the correlation time $\tau_c$ is constantly of the order of the ms, which is simply the inverse of the spectrum frequency span. This {\it microscopic time} $\tau_c$ might simply be the mean flight time of the beads between two collisions. \\
\indent
The measurement described above is re-conducted for several values of the current $I$, at constant value of the vessel acceleration $a=41\,\rm{ms^{-2}}$. The curves shown in fig.~\ref{fig.8} collapse to a unique value $E_c \simeq 1.7\;10^{-7}\,$J. Uncertainties are difficult to evaluate, but it is believed that the dispersion of the curves is mainly due to the fitting process of the asymmetry function, and to statistical limitations. \\
The collapse of the curves in fig.~\ref{fig.8}, obtained at different $I$, can be interpreted as the following. Within a certain range of current amplitudes, $E_c$ does not depend on the blade+motor system and its excitation. It is therefore a characteristic of the granular gas solely. \\
In other words, this externally driven harmonic oscillator actually qualifies as a probe. 
  
\noindent
The range of 'acceptable' currents is however bounded. If $I$ is too large, the fluctuation rate is small, angular velocity follows the torque, little negative events are observed: $E_c$ can hardly be measured. If $I$ is too small, the histograms are almost even, depreciating the resolution on $E_c$. For this reason, the time-lag on which $E_c$ is constant and well defined is smaller for larger currents. This also can be seen in fig.~\ref{fig.8}.\\\\
\indent
It has been shown so far that $E_c$ is an energy characteristic of the granular gas, not of the probing, and how to measure it. No simple argument could be found to account for the actual numerical values of $E_c$. The following steps are to relate $E_c$ to some parameters of the NESS granular gas itself. \\

\begin{figure}[h!]
  \begin{center}
\onefigure[width=8.8cm, height=5.6cm]{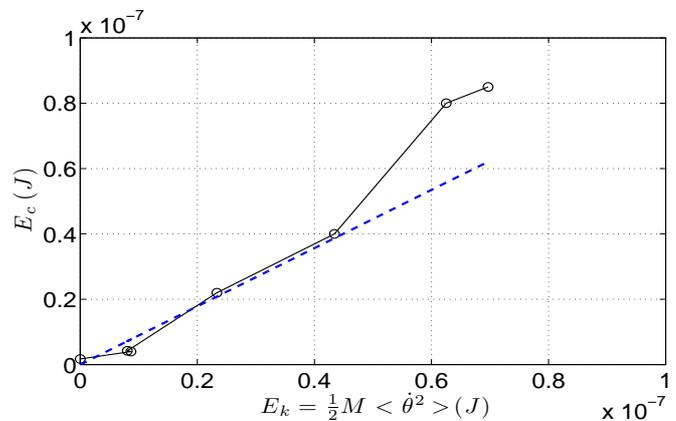}
  \end{center}
\caption{$E_c$ is plotted against the kinetic energy of the blade+rotor system: $E_k={\frac{1}{2}}M <\dot\theta^2 >$. Dashed line is the best linear fit through $0$. 
The acceleration of the vessel is varied between $41\,\rm{ms}^{-2}$ and $59\,\rm{ms}^{-2}$, at constant frequency $f_{\rm{exc.}}=40 \,$Hz, $I=200\, \mu$A unchanged.}
\label{fig.9}
\end{figure}

\indent
Varying the power supply of the shaker $P_{\rm{exc.}}$, one easily vary the acceleration of the vessel and therefore the agitation of the granular gas. A calibration of the motor is performed, giving the coefficient $\alpha= e/{\dot \theta}\simeq0.0029$ with a good accuracy. The calibration of the moment of inertia gives $M=3.2\;10^{-7}\rm{\,kg\,m^2}$, with an uncertainty of about $15\%$. The kinetic energy of the blade+rotor system $E_k=\frac{1}{2}M <\dot \theta^2>$ can be calculated, from the variance of the induced voltage $<e^2>$, recorded with no excitation. In fig.~\ref{fig.9}, $E_c$ is plotted against $E_k$ for several values of the acceleration. A linear fitting, forced at 0, gives a proportionality coefficient of 0.89. Within the uncertainty, $E_c=E_k$. The characteristic energy of the granular gas may be interpreted as the kinetic energy of the probe coupled to the granular gas. \\

\indent
Another parameter that can easily be varied is the acceleration $a$ imposed by the shaker to the beads. How can, dimensionally, an energy be related to an acceleration ? 
\begin{figure}[h!]
  \begin{center}
\onefigure[width=8.8cm, height=5.4cm]{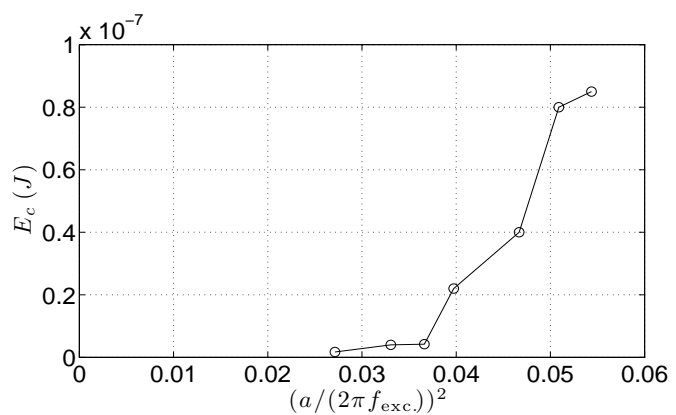}
  \end{center}
\caption{$E_c$ against $(a/(2\pi f_{\rm{exc.}}))^2$, for several values of the acceleration (between $41\,\rm{ms}^{-2}$ and $59\,\rm{ms}^{-2}$). The frequency $f_{\rm{exc.}}=40 \,$Hz, $I=200\, \mu$A unchanged.}
\label{fig.10}
\end{figure}

\noindent
Supposing some sort of {\it thermalisation} between the probe and the granular gas, $E_c$ can be written tentatively in the form: $E_c=E-E_0$, with $E=<mv^2>$. The mass $m$ accounts for a certain mass of beads, and $v$ their velocity. The brackets represent an average on an unspecified number of particles in the vicinity of the blade. $E_0$ is the minimal energy needed to take a large enough number of beads up to the probe. For $E<E_0$, there is no $E_c$ measurable, simply because no beads hit the blade. Still dimensionally, this relation can be rewritten as: $E_c=m' (a/(2\pi f_{\rm{exc.}}))^2-E_0$. No negative $E_c$ nor $E_0$ is to be considered. Apart from small values, data are consistent with this naive picture, at least at the present level of uncertainty, as can be seen in fig.~\ref{fig.10}. Uncertainties on $E_c$ are hard to quantify, but may be large. In these conditions it is vain to further investigate $E_0$ and $m'$, which are system-dependent. 

\section{Conclusion and perspective}
\label{conclusion}
The system presented here is very simple. Nevertheless, it allows very sensitive measurements and gives access to important and topical issues in non-equilibrium statistical mechanics. \\
It consists of a driven harmonic oscillator, coupled as a probe to a granular gas. It is composed by a DC-motor with a blade and a torsion spring on its axis. The granular gas is maintained in a gaseous state by an external power supply $P_{\rm{exc.}}$ from a shaker. It has an extremely large number of degrees of freedom, which guaranties an efficient chaotic loss of memory. It forms a NESS bath, to which the harmonic oscillator {\it thermalises}. The granular gas mimics quite well a real gas, yet with important differences: the number of particles is not extremely large (fluctuations are important), the collisions are dissipative (a power supply is necessary).\\
The DC-motor driving the harmonic oscillator is used as actuator and sensor altogether, to probe the granular gas. The electro-mechanical relations in a DC-motor are such that no calibration is needed to measure the power delivered to the bath. This property is fortunate, as calibration is a source of errors. \\ 
Through this device, a small energy flux $\dot w$ is injected into the granular gas at controlled torque. Note that $<\dot w>$ is of the order of a few $ 10^{-9}\,$W, {\it i.e.} resolution is much less. It is extremely small in such a macroscopic experiment ! \\
The fluctuations of this perturbation power is shown to likely verify the Stationary State Fluctuation Theorem, although the {\it heat bath} is non-thermal. However, it must be noted that this study does not strictly rely on this theorem: the asymmetry function is used to define the output quantity $E_c$. The {\it characteristic energy} does not depend on the imposed torque amplitude within a certain range. One concludes that $E_c$ is characteristic of the granular gas itself, {\it i.e.} this small system is actually a probe. \\
\noindent
For different values of the power $P_{\rm{exc.}}$ supplying the NESS, the characteristic energy $E_c$ is measured, and found equal to the kinetic energy of the swivel device. The dependance in the acceleration is examined. Despite important scattering of the data, a clear tendency to increase is visible, compatible with a quadratic dependance in $a$. This corroborates the intuition. \\
A quantitative characterisation of this relation definitely requires improvements of the experimental set-up and the measurement procedure to measure the asymptotic $E_c$. \\
Considering the FT on a small perturbative energy flux instead of that supplying the NESS is a biased opinion adopted here, which appears fruitful. It allows in principle to measure negative fluctuations of injected power with arbitrarily long statistics, for a granular gas as far as desired from equilibrium. This key point was discussed by Zamponi \cite{zamponi2007} as a limiting benchmark for experiments. \\
One interesting question is whether $E_c$ obtained {\it via} the FT in the present study is the same or different as that obtained using the fluctuations of the power supply $P_{\rm{exc.}}$, that maintains the gas in its NESS. (Like in most of previous studies, for instance \cite{aumaitre2004}). \\
It would be interesting to clarify the physical significance of $E_c$. The characteristic energy is often interpreted, in the context of FT, in terms of an effective (non-equilibrium) temperature: $E_c \equiv k_{\rm{B}}T_{\rm{eff.}}$. 
$E_c$ has been related to the kinetic energy of the probe, but experimental evidence of a relation with the kinetic energy of the gas in the vicinity is still needed to justify thermalisation. Such a measurement is definitely needed to resolve the issue. But such study would require different technics, like PIV. 

Overextending the analogy with a gas, it is worth noticing that the relation obtained in fig.~\ref{fig.9} differs from the prediction of the equipartition theorem, by a factor $2$. \\
In the present experiment, such interpretations deserve caution, as the forcing is non-Gaussian (for short times), the granular gas is intrinsically dissipative and has an irreversible dynamics (the {\it heat bath} is non-thermal). Rigorous interpretation of these observations remains to be done.
\acknowledgments
I gratefully acknowledge S.\,Ciliberto, who has been encouraging and present as an expert adviser, all along this work. Many thanks to A.\,Steinberger, E.\,Bertin, and V.\,Grenard for their help. Many thanks to the colleagues and the students of the \'ENS-Lyon for so many discussions.

\end{document}